\documentclass[twocolumn,showpacs,pra,aps,10pt,superscriptaddress]{revtex4-1}
\usepackage{graphicx}
\usepackage{epstopdf}
\usepackage{amsmath}
\usepackage{amssymb}
\usepackage{epsfig}
\usepackage{amsthm}
\usepackage{color}

\begin{document}

\title{Normalized Stokes operators  for polarization correlations of entangled optical fields}

\author{Marek \.Zukowski}
\author{Wies{\l}aw Laskowski}
\affiliation{Institute of Theoretical Physics and Astrophysics, Faculty of Mathematics, Physics and Informatics, University of Gda\'{n}sk, 80-308 Gda\'{n}sk, Poland}
\author{Marcin Wie\'sniak}

\affiliation{Institute of Informatics, Faculty of Mathematics, Physics and Informatics, University of Gda\'{n}sk, 80-308 Gda\'{n}sk, Poland}



\begin{abstract}
Stokes  parameters  are a standard tool in quantum optics. They involve averaged intensities at exits of polarizers. If the overall measured intensity fluctuates, as e.g. for states with undefined photon numbers, the instances of its increased value contribute more to the parameters. One can introduce normalized quantum  Stokes operators.  Operationally, for a given {\em single} run of the experiment, their values are  differences of measured intensities (or photon numbers) at the two exits of a polarizer divided by their sum. Effects of intensity fluctuations are removed. Switching to normalized Stokes operators   results in  more sensitive entanglement conditions. We also  show a general method of deriving an entanglement indicator for optical fields which use polarization correlations, which starts with any two-qubit entanglement witness. This  allows to vastly expand the family of such indicators.
\end{abstract}

\maketitle
%
%


\section{Introduction}

In 1852 Stokes introduced  his parameters to characterize  polarization of arbitrary states of classical light. The quantum versions are  straightforward application of his ideas. If one assumes for simplicity that the registered   intensity is proportional to the number of photons, the usual quantum Stokes operators read:  
$
{\Sigma}^a_i=a^\dagger_ia_i-a^\dagger_{i\perp}a_{i\perp},
$
where $a_i$ are annihilation operator of photons of polarization $i$,  $a_{i\perp}$ plays the same role for the orthogonal polarization, and the index $i$ denotes three complementary polarization analysis arrangements  (e.g. horizontal-vertical, diagonal-antidiagonal and right-left handed circular). The superscript $a$ denotes the beam (spatial mode).
The fourth Stokes observable  is  the total intensity 
$
{\Sigma}_0^a=\hat{N}^a_{tot}=a^\dagger_ia_i+a^\dagger_{i\perp}a_{i\perp}. 
$
It is invariant with respect to the choice of $i$.

Strictly non-classical optical phenomena are observable in correlations. Especially correlations of polarizations at two or more spatially separated detection stations. Are the above quantum optical definitions of Stokes operators optimal in the domain of correlations? The standard approach is to use for two beams $a$ and $b$ correlation functions
\begin{equation} \label{G}
G(a,i;b,j)=\frac{\langle \Sigma_i^a\Sigma_j^b\rangle}{\langle \Sigma_0^a\rangle \langle\Sigma_0^b\rangle}.
\end{equation}
We shall show that this is not always the optimal. At least for the examples presented below moving to normalized of Stokes observables allows one to detect entanglement  in situations in which the traditional approach fails.

\section{Normalized Stokes observables}

We assume the following measurement procedure defining the normalized Stokes observables. We have a sequence of light pluses, which are equivalently prepared. When $r$-th  pulse  arrives at a detection station  $a$, which consists of a two-output polarization analyzer and pair of detectors, one measures  the photon numbers at each output, respectively $N^a_i(r)$ and $N^a_{i\perp}(r)$.  The  value of the normalized Stokes observable $\hat{S}_i^a$  for the $r$-th run is then
\begin{eqnarray}
&S_i^a(r)=\frac{N^a_i(r)-N^a_{i\perp}(r)}{N^{a}(r)},&
\end{eqnarray}
where  $N^{a}(r)=N^{a}_i(r)+N^{a}_{i\perp}(r)$).
Additionally, we postulate that whenever $N^{a}(r)=0$, we put $S_i^a(r)=0$.
We also introduce $\langle S^a_0\rangle$ as the frequency of runs in which $N^a(r)\neq 0$. Note that operational meaning of the traditional approach is that  we {\em separately} average, over all runs of the experiment, $N^a_i(r)-N^a_{i\perp}(r)$ to get $\langle\Sigma^a_i\rangle$, and $N^a(r)$ to get $\langle \Sigma_0^a \rangle $. The usual normalization of Stokes parameters is via
${\langle\Sigma_i^a\rangle}/{\langle\Sigma_0^a\rangle}$.



The normalized Stokes operators are of little practical value if one considers just one detection station observing polarization effects. E.g.  for light of undefined photon numbers  a possible   degree of polarization defined as $p'=\frac{1}{ \langle S_0^a\rangle}\sqrt{\sum_{i=1}^3 {\langle\hat{S}_i^a\rangle}^2}$ usually gives different values than the usual definition.  If the state is an eigenstate of $\hat{N}^a_{tot}$ then the degrees of polarization are identical.
However, as we shall show, in case of some important entangled states of light,  if one observes polarization correlation at two detection stations, and uses 
$
\langle S_i^aS_j^b\rangle,
$ together with
$\langle S_0^a\rangle$ and $\langle S_0^b\rangle$,
instead of (\ref{G}),
 one can observe effects indicating entanglement much more clearly. For example, we shall formulate a modification of the widely used (necessary) separability condition
of  Ref. \cite{BOUW}:
 \begin{eqnarray}\label{OLD}
&\sum_{i}\langle {\Sigma_i^a}+ {\Sigma_i^b}\rangle_{sep}^2\geq 2\langle \hat{N}^a_{tot}+\hat{N}^b_{tot}\rangle_{sep},&
\end{eqnarray}
where $\langle...\rangle_{sep}$ denotes an average over a separable state.


We have a highly developed theory of entanglement of systems described by finite dimensional Hilbert spaces, see e.g. \cite{HHH}.  Still we  search for entanglement conditions for infinite dimensional systems. We shall show that the notion of normalized Stokes operators  allows us to re-formulate any entanglement witness for two-qubits, like those in \cite{HHH},  into entanglement indicators involving polarization measurements for quantum 
optical fields. Further generalizations are possible.

To the best of our knowledge the normalized  Stokes observables used here cannot be found in the literature.
E.g.,  a recent extensive discussion of proposals for degree of polarization of quantum fields \cite{BJORK} does not cover the ideas presented here.  The unconventional definition of the degree of polarization of Luis 
\cite{LUIS}  is based on different concepts and more involved measurement techniques.

Below,  we shall use the number operator $\hat{n}_i=a_i^\dagger a_i$ as our model for intensity observable. However, obvious generalizations of our formalism to other models \cite{MANDEL-WOLF} exist.


\section{Mathematical formulation}

In the quantum optical formalism the normalized Stokes observables  read:
\begin{equation}\label{STOKES-0-NV}
\hat{S}_i^a=\Pi_a\frac{a^\dagger_ia_i-a^\dagger_{i\perp}a_{i\perp}} {\hat{N}_{tot}} \Pi_a.
\end{equation}
We explain  notation below,  while addressing the most important technical  features of the formula.
In order to avoid problems with vacuum components of states, which give zero in the denominator,  $\hat{S}_i^a$ is formulated in such a way so that it acts only in the non-vacuum sector of the Fock space of photons: symbols $\Pi_a$ stand for projectors 
$\hat{I}-|0,0\rangle_{aa}\langle0,0|$, where  $|0,0\rangle_{a}$ is the vacuum state of the two polarization modes of beam $a$  satisfying $a_i|0,0\rangle_{a}=a_{i\perp}|0,0\rangle_{a} =0.$ We also introduce $\langle\hat{S}_0^a\rangle={\rm Tr}\big[\Pi_a\varrho\big], $ which is the probability of a non-vacuum event. For more mathematical properties of the modified Stokes operators see Appendix \ref{app-math}.


The numerator in the definitions can be put as
$
A^{\dagger}\sigma_i A,
$ where $\sigma_i$ is a Pauli matrix, and $A^\dagger$ is a row matrix $[a^\dagger_H \hspace{2mm} a^\dagger_V]$, while $A$ is its ``column Hermitian conjugate" involving the annihilation operators.
{\em Any} Pauli operator is represented by $\vec{m}\cdot \vec{\sigma}$, where $\vec{m}$ is a unit real vector, and $ \vec{\sigma}$ is a `vector' built out of three  Pauli matrices: $(\sigma_1,\sigma_2,\sigma_3)$. Thus the normalized Stokes operator for any elliptic polarization, associated with the vector $\vec{m}$, reads:
$
\label{STOKES-2}
\vec{m}\cdot\hat{\vec{S}}^a=\Pi_a\frac{A^{\dagger}\vec{m}\cdot\vec{\sigma}A} {\hat{N}_{tot}} \Pi_a.
$
Obviously,  for all $\vec{m}$, one has $|\langle\vec{m}\cdot\hat{\vec{S}}^a\rangle|\leq\langle\hat{S}^a_0\rangle$, and thus  $|\vec{m}\cdot \langle\hat{\vec{S}}^a\rangle|\leq \langle\hat{S}_0^a\rangle$, where $\langle\hat{\vec{S}}^a\rangle$ is a Stokes vector built out of the three components $\langle \hat{S}_i^a\rangle$. The  inequality holds for {\em any unit} $\vec{m}$.  By choosing the $\vec{m}$ which is parallel to $ \langle\hat{\vec{S}}^a\rangle$, one gets  an important  property 
\begin{equation}\label{BLOCH}
\sum_{i=1}^3 {\langle\hat{S}_i^a\rangle}^2 \leq \langle S_0^a\rangle^2\leq 1 .
\end{equation}

Note  that the definition (\ref{STOKES-0-NV}), introduces operators of a completely different nature than the  {\em pseudo-spin} ones \cite{CHEN}. The pseudo-spin operators have as their spectrum just $\pm1$, while the normalized Stokes operators  (\ref{STOKES-0-NV}) have a spectrum, which covers all {\em rational} numbers between $1$ and $-1$. E.g. the $z$ component of pseudo-spin is $(-1)^{\hat{n}}$, where $\hat{n}$ is the number of photons operator for the given  mode.  While one missing photon completely flips the value of the pseudo-spin, in the case of observables  (\ref{STOKES-0-NV}), for higher photon numbers, the value does not change much. 

\section{ Better entanglement conditions: example}

We shall formulate an analog of the separability condition of Ref. \cite{BOUW} for normalized Stokes observables. 
As in Ref.  \cite{BOUW}, as our example of an optical state shall consider  the four mode  squeezed vacuum
\begin{eqnarray}
\label{BSV}
|BSV\rangle=\frac{1}{\cosh^2{\Gamma}}\sum_{n=0}^{\infty} \sqrt{n+1} \tanh^{n}{\Gamma} |\psi^{(n)}_- \rangle.
\end{eqnarray}
The $2n$ photon singlets in (\ref{BSV}) are given by
\begin{eqnarray}
&|\psi^{(n)}_-\rangle&\nonumber\\
&=\frac{1}{\sqrt{n+1}} \sum_{m=0}^n (-1)^m|n-m\rangle_{a_H} |m\rangle_{a_V} |m \rangle_{b_H} |n-m\rangle_{b_V},&\nonumber\\
\end{eqnarray}
where $a$ and $b$ refer to the two directions along which the photon pairs are emitted,  $H/V$ denote horizontal/vertical polarization, and $\Gamma$ represents an amplification gain, which is proportional  to the strength of the pump and the coupling. The state  represents  (strongly) driven type II parametric down conversion process \cite{BOUW, PAN}.

\subsection{Separability condition based on EPR correlations}

An analogue of the separability condition of Ref. \cite{BOUW}, see inequality (\ref{OLD}), for standard Stokes operators can be formulated by employing the intuition that for the two photon singlet, and also  for four mode bright squeezed vacuum  state (a generalized singlet, see e.g. \cite{ROSOLEK} ) one has 
\begin{eqnarray}\label{EPR-1}
&\sum_{i}\langle(\hat{S}_i^a+\hat{S}_i^b)^2\rangle=0.&
\end{eqnarray}
This EPR condition can also be put in a more sophisticated form  
which is a reformulation of the condition given in Iskhakov et al.\cite{MASZA}, but we shall not discuss this here.
We shall show below  that { for {\em no} separable state the expression (\ref{EPR-1}) can be zero}.


 One has 
\begin{eqnarray}\label{EPR-5}
&\sum_{i}\langle(\hat{S}_i^a+\hat{S}_i^b)^2\rangle_{sep}&\nonumber\\
&=\sum_{i}\langle\hat{S}_i^{a2}+\hat{S}_i^{b2} + 2 \hat{S}_i^a\hat{S}_i^b)\rangle_{sep}.&
\end{eqnarray}
Recalling the well known formula for the usual Stokes operators (see e.g. Klyshko \cite{KLYSHKO}): 
\begin{eqnarray}
&\sum_{i}{\hat{\Sigma}_i^{a2}}= \hat{N}_{tot}(\hat{N}_{tot}+2),&
\end{eqnarray}
one can  find its equivalent for the new Stokes operators
\begin{eqnarray}
&\sum_{i} \hat{S}^{a2}_i= \Pi_a+2\Pi_a\frac{1}{\hat{N}^a_{tot}}\Pi_a.&
\end{eqnarray}
Therefore,  the values of the first two terms of the RHS of (\ref{EPR-5}) are $ \langle\hat{S}_0^a\rangle_{sep}+2\langle \Pi_a\frac{1}{\hat{N}^a_{tot}}\Pi_a\rangle_{sep}$ and  $ \langle\hat{S}_0^b\rangle_{sep}+2\langle\Pi_b \frac{1}{\hat{N}^b_{tot}}\Pi_b\rangle_{sep}$. The lowest possible value of $\sum_i\langle\hat{S}_i^a\hat{S}_i^b\rangle_{sep}$ can established by the following observations. 
Note that,  the decomposition of a separable state into a probabilistic mixture of pure states is given by $ \sum_\lambda p_\lambda \varrho^{a}(\lambda)\varrho ^{b}(\lambda)$. Each of the local states $\rho^k(\lambda)$ is endowed with normalized  
Stokes parameters   $\vec{s}_{k}(\lambda)=Tr[\vec{\hat{S}}^k\varrho^k(\lambda)]$ and  ${s}_{k0}(\lambda)=Tr[{\hat{S}}_0^k
\varrho^k(\lambda)]$,  where $k=a,b$.  
Using the above one gets
 \begin{eqnarray}&\label{ZZZ}\sum_i\langle\hat{S}_i^a\hat{S}_i^b\rangle_{sep}=\sum_\lambda p_\lambda \vec{s}_{a}(\lambda)\cdot \vec{s}_{b}(\lambda).&\end{eqnarray}
 The following holds for any vectors: $2\vec{s}_{a}(\lambda)\cdot \vec{s}_{b}(\lambda)\leq |\vec{s}_{a}(\lambda)|^2+|\vec{s}_{b}(\lambda)|^2$.  This in turn is less than $|\vec{s}_{a}(\lambda)|+|\vec{s}_{b}(\lambda)|$, because the local normalized Stokes  vectors in the expression cannot have norms larger than $1$, see (\ref{BLOCH}). Next we notice that $|\vec{s}_{k}(\lambda)| \leq {s}_{k0}(\lambda)$, and finally that $\langle\hat{S}_0^k\rangle_{sep}=\sum_\lambda p_\lambda {s}_{k0}(\lambda).$ Therefore, we reach 
\begin{eqnarray}
2\min \sum_\lambda p_\lambda \vec{s}_{a}(\lambda)\cdot \vec{s}_{b}(\lambda)\geq-\langle\hat{S}_0^a\rangle_{sep}-\langle\hat{S}_0^b\rangle_{sep}.
\end{eqnarray}
Thus,  a {\em necessary} condition for a state to be separable reads:
\begin{eqnarray}\label{EPR-SEP}
&\sum_{i}\langle(\hat{S}_i^a+\hat{S}_i^b)^2\rangle_{sep}&\nonumber\\
&\geq 2 \big(\langle \Pi_a\frac{1}{\hat{N}^a_{tot}}\Pi_a\rangle_{sep}+\langle \Pi_b \frac{1}{\hat{N}^b_{tot}}\Pi_b\rangle_{sep} \big).&
\end{eqnarray}

\subsection{Comparison with the earlier approach}

In Appendix \ref{app-eff} we show  that the  condition (\ref{EPR-SEP}), in the case of noise modeled by photon losses (non-perfect efficiency of detection),  detects the  entanglement of $|BSV\rangle$  better than the analogue  condition (\ref{OLD}),  Ref. \cite{BOUW}.
No matter what is the gain parameter $\Gamma$, the standard condition (\ref{OLD}) fails to detect the entanglement in  $|BSV\rangle$  for $\eta\leq 1/3$, see Bouwmeester and Simon \cite{BOUW}, while  the new condition still  works for lower efficiencies than $1/3$. The actual threshold $\eta(\Gamma) $ is a {\em decreasing} function of $\Gamma$, which is less than $1/3$ for all $\Gamma>0$. 

This has interesting  ramifications.  
The condition (\ref{EPR-SEP}) allows the following. 
In theory, for perfect detection case, $\eta=1$, one can beam-split both beams, $a$ and $b$,  in a polarization neutral way, by using three output polarization-neutral beam-splitters (tritters), of the property that  they split the incoming beams into three beams of equal (average) intensities. If we now place  at the exits of the local beam-splitters three polarization measurement stations,  set to measure simultaneously three complementary polarizations (e.g. horizontal-vertical, diagonal-antidiagonal, left-right circular), the conditions  (\ref{OLD})
 would not be capable to detect entanglement of $|BSV\rangle$. However, condition  (\ref{EPR-SEP}) would still detect entanglement, because the pairs of identical polarization measurements devices, one at side $a$, the other at $b$, would give correlations as if we had an experiment without the tritter, but with detection efficiency $\eta=1/3$. Thus while the old condition obeys the standard `complementarity rule of thumb', the new one does not. Of course the reason, for   circumventing the  complementarity rule in the second case, is that in the case of $|BSV\rangle$ we do not have defined photon numbers, and the state has components with arbitrarily high photon numbers, $|\psi^{(n)}_-\rangle$. Strict polarization complementarity rule  works in the case of condition  (\ref{EPR-SEP}) only for the component of BSV with  one photon in beam $a$ and one photon in beam $b$, that is for the singlet  $|\psi^{(1)}_-\rangle$ .  

The above remarks hold also for the singlets  $|\psi^{(n)}_-\rangle$ themselves, for $n\geq 2.$ For states of fixed total photon number, like $|\psi^{(n)}_-\rangle$, and perfect detection, the two conditions are fully equivalent. However, surprisingly, if one introduces the detection losses, the condition (\ref{EPR-SEP}) performs much better than (\ref{OLD}). This is the more pronounced the higher is $n$. In the limit of $n\rightarrow \infty$, the threshold efficiency for condition (\ref{EPR-SEP}) approaches $0$, while for (\ref{OLD}) it stays put at $1/3$ (see Appendix \ref{app-eff}).

\section{Constructing polarization entanglement indicators for quantum optical fields}

One can map entanglement conditions for qubits, for a review see  $4\times$Horodecki \cite{HHH}, into entanglement indicators for optical fields employing the new polarization parameters. 
We present this for two beam situations. Generalizations are obvious.

{\em The map.}
Take an entanglement witness, $\hat{W}$, or any other indicator of two qubit non-separability. Expand it in terms of local Pauli operators. This is always possible as Pauli observables form the basis in the linear space of all one-qubit observables. We get $\hat{W}={W}(\sigma^a_\mu, \sigma^b_\nu)$, where $\mu,\nu=0,1,2,3$, and $a,b$ now denote the qubits. Finally we  make a replacement:
$
\sigma_i ^k\rightarrow \hat{S}_i^k.
$
and 
$
\sigma_0^k \rightarrow \hat{S}_0^k,
$
to get a quantum optical witness $\hat{W}_{QO}= W({\hat{S}^a_\mu, \hat{S}^a_\nu}).$ Next one has to find the upper or lower bound for this operator in the case of separable states of optical fields,  that is $B_{min}\leq \langle \hat{W}_{QO}\rangle_{sep}$,
 or $B_{max}\geq \langle  \hat{W}_{QO}\rangle_{sep}$, one of which gives the necessary condition for separability.

To illustrate this,
let us take the  condition for separability of two-qubit states derived by Yu et al.\cite{PAN2002}. We choose this example because of its generality. The condition of Yu et al. is equivalent to the partial transposition condition (PPT), which is a sufficient and necessary separability condition for two qubit states. It reads
\begin{eqnarray}\label{PAN1}
&\langle\sigma^a_x\sigma^b_x + \sigma^a_y\sigma^b_y\rangle^2 + \langle\sigma^a_z\sigma^b_0 +\sigma^a_0\sigma^b_z\rangle^2& \nonumber \\
&\leq \langle \sigma^a_0\sigma^b_0 +\sigma^a_z\sigma^b_z\rangle^2,&  
\end{eqnarray}
for any choice of orthogonal directions $\vec{x},\vec{y},\vec{z}$. 
This is mapped to
\begin{eqnarray}\label{PAN2B}
&\frac{1}{\langle\hat{S}^a_0\hat{S}^b_0\rangle }\left(\sqrt {\langle\hat{S}^a_x\hat{S}^b_x + \hat{S}^a_y\hat{S}^b_y\rangle{^2} + \langle\hat{S}^a_z\hat{S}^b_0 +\hat{S}^a_0\hat{S}^b_z\rangle{^2}}\right.& \nonumber \\
&-\left.\langle \hat{S}^a_z\hat{S}^b_z\rangle\right)\leq 1,& 
\end{eqnarray}
where we use the convention that $i=x,y,z$, 
As it cannot be for sure a necessary {\em and sufficient} condition for separability of the quantum optical states, we shall now give only the proof of its necessity (i.e., that a violation of this condition indicates entanglement).

The inequality (\ref{PAN1})  holds also for  any  pure product  state of two qubits. Thus the Bloch vectors  of the two qubits, $\vec{b}^a$ and $\vec{b}^b$ must satisfy:
\begin{eqnarray}\label{PAN3}
&0\leq 1 +b^a_z b^b_z-  \sqrt{(b^a_x b^b_x + b^a_y b^b_y)^2 + (b^a_z +b^b_z)^2}.& 
\end{eqnarray}
This can be linearized, as for any  $\alpha$
\begin{eqnarray}\label{PAN4}
&0\leq 1 +b^a_z b^b_z +\cos{\alpha}(b^a_x b^b_x + b^a_y b^b_y) +\sin{\alpha} (b^a_z +b^b_z).& \nonumber\\
\end{eqnarray}
Next notice, that the above inequality holds   for Bloch vectors of products of mixed states of two qubits. Thus one can have $|\vec{b}^k|\leq 1$. 
Therefore,  if one introduces two numbers $b^a_0$, and $b^b_0$, one has such that $|\vec{b}^k|\leq b^k_0\leq 1$, one has
\begin{eqnarray}\label{PAN4B}
&0\leq b^a_0b^b_0 +b^a_z b^b_z +\cos{\alpha}(b^a_x b^b_x + b^a_y b^b_y) +\sin{\alpha} (b^a_zb^b_0  +b^a_0 b^b_z).&\nonumber  \\ 
\end{eqnarray}
Ineq. (\ref{PAN4B}) can be used for the components of  vectors $\vec{s}_k(\lambda)$, and parameters $s_{0k}(\lambda)$ introduced  earlier, which are  the   Stokes-like parameters   for product states of light in beams $a$ and $b$, which enter the convex expansion of a given separable state into product states. We have 
\begin{eqnarray}\label{PAN5}
&0\leq s_{a0}s_{b0} +s_{az} s_{bz}+\cos{\alpha}(s_{ax} s_{bx} + s_{ay} s_{by}) &\nonumber  \\& +\sin{\alpha} (s_{az}s_{b0} +s_{a0}s_{bz}), & 
\end{eqnarray}
where the symbols $(\lambda)$ were dropped. After averaging over probability $p_\lambda$, and using the Cauchy inequality for the terms with trigonometric functions one gets (\ref{PAN2B}). QED.

With such a techniques one can derive necessary conditions for separability based on any other two-qubit entanglement criterion.

\subsection{Separability conditions with standard quantum Stokes parameters}

Note, that such conditions have their equivalents in the traditional approach to Stokes parameters. In such a case product states  $\rho^k{(\lambda)}$ are endowed with Stokes vectors of arbitrary lengths. Let us denote their components by $z_i^a(\lambda)$ and $z_j^a(\lambda)$. It is obvious that the following algebraic identity holds ($\lambda$'s are again dropped):
\begin{eqnarray}\label{PAN7}
&0\leq ||\vec{z}^a||||\vec{z}^b|| +z^a_z z^b_z&  \nonumber \\&-  \sqrt{(z^a_x z^b_x + z^a_y z^b_y)^2 + (z^a_z||\vec{z}^b|| +||\vec{z}^a||z^b_z)^2}.&
\end{eqnarray}
For any  $\rho^k{(\lambda)}$, one has $\langle \hat{n}_{tot}^k\rangle_{(\lambda)}=Tr[N_{tot}^k\rho^k{(\lambda)}]\geq ||\vec{z}^k(\lambda)||$.  Thus in inequality  (\ref{PAN7}), one can  replace $||\vec{z}^k(\lambda)||$ by $\langle \hat{n}_{tot}^k\rangle_{(\lambda)} $, just as it was done in (\ref{PAN5}). Upon convex summation over the probabilities of the   product states in the separable state, one reaches the following separability condition with traditional Stokes operators
\begin{eqnarray}\label{PAN8}
&\frac{1}{ \langle  \hat{N}_{tot}^a\hat{N}_{tot}^b\rangle }\left(\sqrt{\langle\hat{\Sigma}^a_x\hat{\Sigma}^b_x + \hat{\Sigma}^a_y\hat{\Sigma}^b_y\rangle{^2} + \langle\hat{\Sigma}^a_z \hat{N}_{tot}^b  + \hat{N}_{tot}^a\hat{\Sigma}^b_z\rangle{^2}}\right.& \nonumber \\
&-\left.\langle\hat{\Sigma}^a_z\hat{\Sigma}^b_z\rangle\right)\leq 1.& 
\end{eqnarray}

\subsection{Comparison of conditions (\ref{PAN2B}) and (\ref{PAN8})}
Fig. 1 shows
the strength of violation
of the separability conditions (\ref{PAN8}) and (\ref{PAN2B}) by the bright squeezed vacuum. Normalized Stokes observables outperform the traditional ones for all finite $\Gamma$. This signals a better noise tolerance (see Appendix \ref{app-bsv} for detailed calculations). 
\begin{figure}
\centering
 \includegraphics[width=0.45\textwidth]{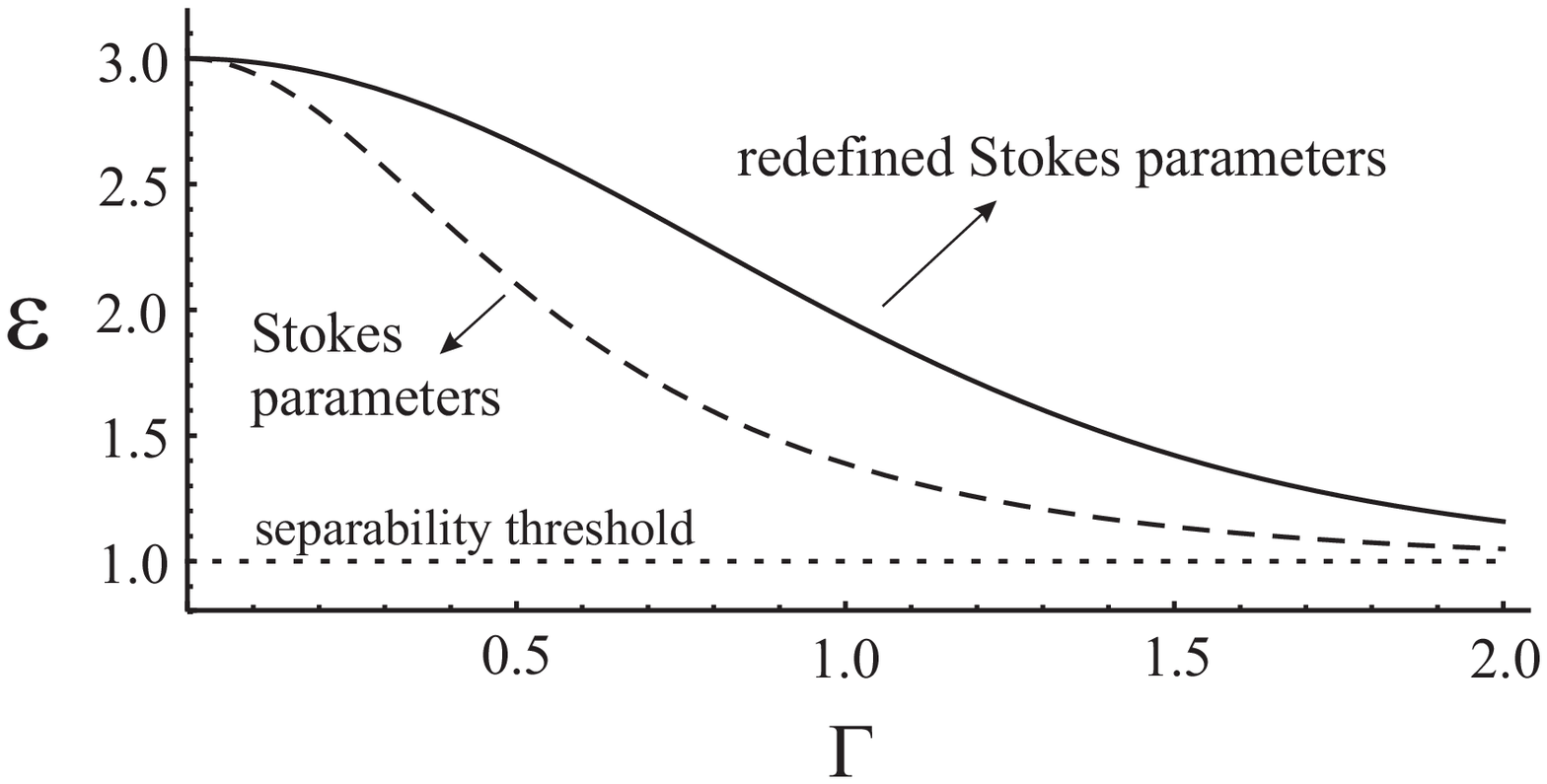}
\caption{\label{fig_bsv} Comparison of entanglement criteria for the BSV state.  The criteria are based on different definitions of the Stokes observables:  traditional  (\ref{PAN8}) and normalized (\ref{PAN2B}). $\Gamma$ is the amplification gain, see  (\ref{BSV}). The symbol $\cal{E}$ stands for the value of the left hand sides. Above level of $1.0$ we detect entanglement. The gap between the two curves indicates more robust violations of separability the the condition based on the normalized Stokes observables. This implies higher noise resistance. }
\end{figure}
A different example, based on the approach presented here, less general and less sensitive to entanglement, but still beating its analogue expressed in terms of standard Stokes operations can be found in \cite{SCRIPTA}.


\section{Final remarks}

The redefined approach to polarization correlations  of quantum states of light with undefined total photon number, allows us to see violations of separability, in experiments using polarization measurements, in situations in case of which more traditional approaches fail to detect entanglement. 

The intuitive reason for this is that in the traditional approach the average total intensities  are  used  to ``normalize" the correlation function $\langle\hat{\Sigma}_i^a\hat{\Sigma}_j^b\rangle$, while in our approach,  we have a normalized polarization measurement in each run.  We use averages of  correlations of ``polarization events" with normalized read-out values of the Stokes parameters. They are totally independent of the measured intensity (fluctuating from run to run). Photon's polarization is an observable which is independent of its momentum and  energy, thus our re-normalization is in tune with this intuitive aspect.

Run-by-run measurements of total intensity and polarization parameters are possible and in fact performed in the labs \cite{MASZA}. Most importantly, measurements of normalized Stokes observables do {\em not} require any special new techniques.
Just as for correlations of the standard Stokes observables, what one needs to register in each experimental run $r$ are $N_i^a(r)$, $N_{i\perp}^a(r)$,  $N_i^b(r)$, and $N_{i\perp}^b(r)$.

 Our results show that one can detect entanglement of optical fields, using only polarization measurements, for a significantly broader families of states, than in the case of the traditional approach.  In a separate work \cite{ZWL} we show that the method can be tailored in such a way, so that one can construct Bell inequalities for optical fields, based only on the assumptions of realism, locality and `freedom'. Such (fully) device independent entanglement conditions are, surprisingly, violated by a wider class of states than standard Bell inequalities \cite{REID} involving intensities (and requiring additional assumptions).

The approach can be extended to multi-party situations, and  beyond polarization measurements, see our e-prints \cite{ARXIV} and forthcoming manuscripts.

\section{Acknowledgments}

The work is a part of  EU grant BRISQ2.  The work was additionally subsidized
form funds for science of MNiSW for years 2012-2015 approved
for international co-financed project BRISQ2.  M\.Z and WL were supported by TEAM project of FNP. MZ acknowledges FNP-DFG Copernicus Award and discussions with profs. Maria Chekhova and Harald Weinfurter. WL is supported by NCN Grant No. 2014/14/M/ST2/00818.  MW is supported by NCN Grant No. 2013/11/D/ST2/02638.


\appendix

\section{Mathematical properties of the modified Stokes operators}

\label{app-math}

The normalized Stokes operators written up using  the quantum optical formalism read:
\begin{equation}\label{app-STOKES-0-NV}
\hat{S}_i^a=\Pi_a\frac{a^\dagger_ia_i-a^\dagger_{i\perp}a_{i\perp}} {\hat{N}_{tot}} \Pi_a.
\end{equation}
The basic properties of the operators were explained in the main text. Here we show their
 other properties.
Please notice that operators $a^\dagger_ia_i$ and $a^\dagger_{i\perp}a_{i\perp}$ as well as  $\hat{N}_{tot}= a^\dagger_ia_i+a^\dagger_{i\perp}a_{i\perp}$ obviously all commute with each other. But so does the projector  $\Pi_a
=\hat{I}-|0,0\rangle_{aa}\langle0,0|$, it commutes with all of them. The joint eigen-basis for all these self-adjoint  operators
is the Fock basis, with states $|n_i, n_{i \perp}\rangle_a$, where $n_i$ and $ n_{i \perp}$ are non-negative integers, with the notation defined by the eigenvalues
\begin{equation}
a^\dagger_ia_ia^\dagger_{i\perp}a_{i\perp}|n_i, n_{i \perp}\rangle_a=n_in_{i \perp}|n_i, n_{i \perp}\rangle_a.
\end{equation}
 Thus, as all its constituents are self adjoint linear operators, so is $\hat{S}_i^a$. 
Notice that as mixed states are described by density operators, which are also linear and self-adjoint,
for any convex combination of any two such states  $\hat{\rho}_1$ and $\hat{\rho}_2$, given by $p_1\hat{\rho}_1+p_2\hat{\rho}_2$, with $p_1$ and $p_2$ positive and $p_1+p_2=1$, one has the usual algebraic property that  $\hat{S}_i^a(p_1\hat{\rho}_1+p_2\hat{\rho}_2)=
p_1\hat{S}_i^a\hat{\rho}_1+ p_2 \hat{S}_i^a\hat{\rho}_2.$
Therefore all the general results given in the main text (that is, the inequalities forming conditions for separability) apply both to pure and mixed states.  We have chosen as our working example the pure bright squeezed vacuum state only because of its importance in quantum optics.

\section{Bright squeezed vacuum}

\label{app-bsv}

The  (four-mode, bright) squeezed vacuum state is given by the following formula 
\begin{eqnarray}
\label{app-BSV}
|BSV\rangle=\frac{1}{\cosh^2{\Gamma}}\sum_{n=0}^{\infty} \sqrt{n+1} \tanh^{n}{\Gamma} |\psi^{(n)}_- \rangle,
\end{eqnarray}
where
\begin{eqnarray}\label{app-PSI}
&|\psi^{(n)}_-\rangle&\nonumber\\
&=\frac{1}{\sqrt{n+1}} \sum_{m=0}^n (-1)^m|n-m\rangle_{a_H} |m\rangle_{a_V} |m \rangle_{b_H} |n-m\rangle_{b_V}.&\nonumber\\
\end{eqnarray}
The state is endowed with perfect EPR correlations, we have
\begin{equation}\label{app-EPR-1}
\sum_{i}\langle(\hat{S}_i^a+\hat{S}_i^b)^2\rangle=\sum_{i}\langle(\hat{\Sigma}_i^a+\hat{\Sigma}_i^b)^2\rangle=0,
\end{equation}
for all values of $\Gamma$.  The state is a result of action of type II parametric down conversion Hamiltonian, proportional to 
$a^\dagger_Hb^\dagger_V-a^\dagger_V b^\dagger_H+h.c.$ on the initial state, which is vacuum in all modes. The gain parameter $\Gamma$ depends on the pump power, interaction time (essentially, duration of the pump pulse), and the coupling.

For $|BSV\rangle$ non-vanishing correlation tensor elements, defined by $T'_{ij}=\langle \hat{S}^a_i\hat{S}^b_j\rangle/\langle \hat{S}^b_0\hat{S}^b_0\rangle$, read
\begin{align}
&T'_{11}= T'_{22} = T'_{33}& \\
&=\frac{16 \ln(1/\cosh^2{\Gamma})- \cosh{4 \Gamma}- 12 \cosh{2 \Gamma}+13}{ 12 \sinh^2{\Gamma} (3 + \cosh{2 \Gamma})}.& \nonumber
\end{align}
while non-zero $T_{ij}=\langle \hat{S}^a_i\hat{S}^b_j\rangle$ are given by
\begin{align}
&{T}_{11}= T_{22} = T_{33}& \\
&=\frac{1}{3} \left(\frac{2 \ln(1/\cosh^2{\Gamma}) -\cosh{2\Gamma} +2}{\cosh^4 \Gamma} -1 \right).& \nonumber 
\end{align}
For the traditional Stokes parameters the correlation tensor reads $\Theta_{ij}=\langle \hat{\Sigma}^a_i\hat{\Sigma}^b_j\rangle/\langle \hat{\Sigma}^b_0\hat{\Sigma}^b_0\rangle$, and we have 
\begin{align}
{\Theta}_{11}= \Theta_{22} = \Theta_{33} = \frac{2 \cosh^2 \Gamma}{1-3 \cosh 2\Gamma}.
\end{align}
We also have $\langle \hat{S}^a_i\hat{S}^b_0\rangle= \langle \hat{\Sigma}^a_i\hat{\Sigma}^b_0\rangle=0$, and $\langle \hat{S}^a_0\hat{S}^b_i\rangle= \langle \hat{\Sigma}^a_0\hat{\Sigma}^b_i\rangle=0.$
The above formulas are used to get the curves of Fig. 1.

\subsection{Calculation technique}


In the main text we  compare the strength of separability conditions 
\begin{eqnarray}\label{app-OLD}
\sum_{i}\langle {(\Sigma_i^a}+ {\Sigma_i^b})^2\rangle_{sep}\geq 2\langle \hat{N}^a_{tot}+\hat{N}^b_{tot}\rangle_{sep}
\end{eqnarray} 
and 
\begin{eqnarray}\label{app-EPR-SEP}
&\sum_{i}\langle(\hat{S}_i^a+\hat{S}_i^b)^2\rangle_{sep}&\nonumber\\
&\geq 2 \big(\langle \Pi_a\frac{1}{\hat{N}^a_{tot}}\Pi_a\rangle_{sep}+\langle \Pi_b \frac{1}{\hat{N}^b_{tot}}\Pi_b\rangle_{sep} \big)&\\ \nonumber
\end{eqnarray}
in the case of losses (see the main text for explanation of the notation). 

We perform our calculations using the properties of the  $2n$ photon singlets $\psi_-^{n}$, formula (\ref{app-PSI}), which are components of the bright squeezed vacuum state  (\ref{app-BSV}). This is possible because of  the following observation. The bright squeezed vacuum is a rotationally invariant state, and such are also all $\psi_-^{n}$. Therefore, for each  $\psi_-^{n}$  the three squares on the left hand sides of  (\ref{app-OLD}), and also of (\ref{app-EPR-SEP}),
 will be equal to each other (in both old and new inequalities).  Hence, when considering the left hand sides, it is enough to consider only a square of one component, e.g. in (\ref{app-EPR-SEP}) just $(\hat{S}_i^a+\hat{S}_i^b)^2$,  and multiply the result  it by three.
 This greatly simplifies the calculations.
Further, as none of the operators used in (\ref{app-OLD}) and (\ref{app-EPR-SEP}) changes the total photon number the averages for these conditions can be calculated as a sum of averages for the component singlets $\psi_-^{n}$. Thus effectively we have e.g. 
\begin{eqnarray}\label{app-OLD-1}
&\langle {(\Sigma_i^a}+ {\Sigma_i^b})^2\rangle_{BSV}&\nonumber\\ &=\sum_{n=0}^\infty |C(n, \Gamma)|^2 \langle {\psi_-^{n}} |{(\Sigma_i^a}+ {\Sigma_i^b})^2|{\psi_-^{n}}\rangle,&
\end{eqnarray} 
where $ C(n, \Gamma)$are the expansion  coefficients in the formula for the squeezed vacuum (\ref{app-BSV})
 This also apples to the RHSs of the criteria,  as all operators there do not change the total number of photons in each component of BSV, $\psi_-^{n}$.

Similar remarks apply to calculations of the correlation tensor elements.

\section{Squeezed vacuum with losses: violations of separability conditions}

\label{app-eff}

Here we study to what an extent a noise, due to losses, affects violations of the conditions (\ref{app-OLD}) (\ref{app-EPR-SEP}) by polarization correlations generated by the squeezed vacuum.

For simplicity we shall assume that only our detectors are inefficient (no losses in transmission channels). This will be modeled in the usual way, by a perfect photon-number resolving detector, which however  reports a registered photon only with a probability (efficiency) $\eta<1$.

We shall show that for the condition (\ref{app-OLD}) the threshold efficiency is $\eta=\frac{1}{3}$. This agrees with the value given in \cite{BOUW}. This threshold value does not change with the gain parameter $\Gamma$. In contrast  the threshold efficiency is lower for the condition (\ref{app-EPR-SEP}). The critical efficiency is less than $1/3$ for all non-zero $\Gamma$'s, and is a decreasing function of $\Gamma$.

We perform our analysis using the properties of the  $2n$ photon singlets $\psi_-^{n}$, formula (\ref{app-PSI}). Losses, within our model do not break the rotational  invariance. Hence, when considering the left hand sides, it is enough to consider only a square of one component  and multiply it by three.
 Please note that, 
 our approach  is to  assume that at each side the true number of photons is detected,  thus as in the perfect efficiency case  in each run we have collapses to 
the  $\psi_-^{n}$ states.

\subsection{Calculation of critical efficiency for the singlets  $\psi_-^{n}$}

Assume that in a run of the experiment $n^X$ photons, in total, reach the detectors of observer $X$, out of that $n^X_{i,+}$ and $n^X_{i,-}$ ($X=a,b$) in respective modes ($+,-$ denote the two outputs of an analyzer set to distinguish between polarization $i$ and $i_{\perp}$). However, only $m^X_{i,+}\leq n^X_{i,+}$ and $m^X_{i,-}\leq n^X_{i,-}$ are actually registered by each detector.
 The probabilities of  registration numbers are given  by the binomial  distribution. Namely, the probability that we  register $m^X_{\pm}$ photons in a certain mode, given that we should have been seen $n^X_{\pm}$, for the detector efficiency  $\eta$, reads
\begin{equation}
p(m^X_\pm|n^X_\pm,\eta)=\left(\begin{array}{c}n^X_\pm\\m^X_\pm\end{array}\right)\eta^{m^X_\pm}(1-\eta)^{n^X_\pm-m^X_\pm}.
\end{equation}

Let us first analyze the  criterion (\ref{app-OLD}). 
For $\psi_-^{n}$, let us establish the critical $\eta$, such that after losses the inequality is no longer violated, that is we have
\begin{equation}
\label{app-crit1a}
{\rm LHS}_{(old)}^n\geq {\rm RHS}_{(old)}^n,
\end{equation}
where $\rm LHS_{(old)}^n$ denotes the LHS of inequality (\ref{app-OLD}),  and ${\rm RHS}_{(old)}^n$ is the RHS of it, both calculated for $\psi^n_-$ and inefficient detectors.  One has 
\begin{widetext}
\begin{eqnarray}&{\rm LHS}_{(old)}^n
=3\frac{1}{n+1}\sum_{i=0}^{n}\sum_{j,m=0}^i\sum_{k,l=0}^{n-i}p(j|i,\eta)p(k|n-i,\eta)p(l|n-i,\eta)p(m|i,\eta)(\frac{j-k+l-m}{2})^2
=3\frac{n}{2}\eta(1-\eta),&\nonumber\\
\end{eqnarray}
and right hand side reads 
\begin{eqnarray}
{\rm RHS}_{(old)}^n=\frac{1}{n+1}\sum_{i=0}^{n}\sum_{j,m=0}^i\sum_{k,l=0}^{n-i}p(j|i,\eta)p(k|n-i,\eta)p(l|n-i,\eta)p(m|i,\eta)\frac{j+k+l+m}{2}
=\eta n.
\end{eqnarray}
\end{widetext}
It is easy to verify that condition (\ref{app-crit1a}) is satisfied for any $n$,  provided  $\eta\leq \frac{1}{3}$. 

\begin{figure}
\includegraphics[width=0.45\textwidth]{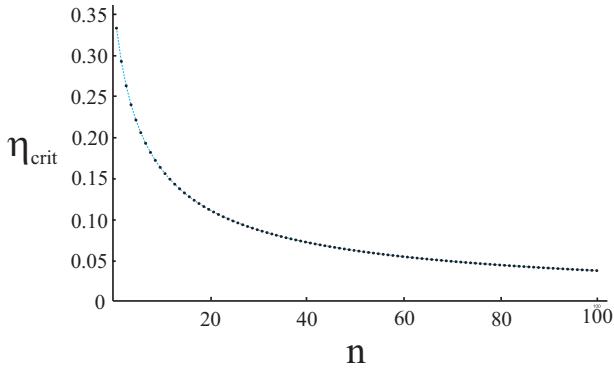}
\caption{\label{fig_eff} The values of the critical efficiency $\eta_{crit}$ for the states $\psi_-^{n}$ with $n \leq 100$.The starting value for $n=1$ is exactly $1/3.$}
\end{figure}

Similar relations for the condition (\ref{app-EPR-SEP}) can be put as follows. If the condition is no longer  violated by $\psi^n_-$ (after the losses) one has  
\begin{equation}
\label{app-crit2a}
{\rm LHS}^n_{(new)}\geq{\rm RHS}^n_{(new)},
\end{equation}
where
\begin{widetext}
\begin{eqnarray}
&{\rm LHS}^n_{(new)}=3\frac{1}{n+1}\sum_{i=0}^{n}\sum_{j,m=0}^i\sum_{k,l=0}^{n-i}p(j|i,\eta)p(k|n-i,\eta)p(l|n-i,\eta)p(m|i,\eta)\left((1-\delta_{j+k})\frac{j-k}{j+k}+(1-\delta_{l+m})\frac{l-m}{l+m}\right)^2,&\nonumber\\
&{\rm RHS}^n_{(new)}=\frac{1}{n+1}\sum_{i=0}^{n}\sum_{j,m=0}^i\sum_{k,l=0}^{n-i}p(j|i,\eta)p(k|n-i,\eta)p(l|n-i,\eta)p(m|i,\eta)\left((1-\delta_{j+k})\frac{2}{j+k}+(1-\delta_{l+m})\frac{2}{l+m}\right).&\nonumber\\
\end{eqnarray}
\end{widetext}
The symbol  $\delta_{k+l}$ denotes the Kronecker delta, with its non-zero value for $k+l=0$. The deltas have to be executed first. Their role is to remove any contribution of terms with no registered photons at each side. We have numerically found $\eta$ saturating Ineq. (\ref{app-crit2a}) for up to $n=100$. The values for low $n$ are given in Fig. \ref{fig_eff}.

The values of $\eta_{crit}$  for the $\psi^n_-$ singlets follow the  function $\eta_{crit} = 1 -  (\frac{2}{n+2})^{1/n}$, at least up to $n=100$. Note that this suggests that for $n\rightarrow \infty$ the critical $\eta$ approaches zero! 
Thus, as  $|BSV\rangle$ is a superposition of states  $\psi_-^{n}$, the critical efficiency to detect entanglement with the condition involving new Stokes parameters is for all values of $\Gamma$ less that $1/3$, and decreases with growing $\Gamma$. Simply, for high $\Gamma$ the terms with higher $n$'s contribute more, this is because of the form of expansion coefficients: $C(n, \Gamma)=  \sqrt{n+1} \frac{ \tanh^{n}{\Gamma}}{\cosh^2{\Gamma}}.
$


\end{document}